\documentclass[12pt]{article}
\usepackage{graphicx,sectsty,amsthm,amssymb,multirow,amsmath}

\usepackage[english,american]{babel}
\usepackage[T1]{fontenc}
\usepackage[ansinew]{inputenc}
\usepackage[pagewise]{lineno}
\usepackage[round, comma,authoryear]{natbib}
\usepackage{epstopdf,bbm,algorithm}
\usepackage{setspace}
\sectionfont{\nohang \centering}

\newcommand{\citepp}[1]{
   (\citeauthor{#1} (\citeyear{#1}))
}
\newcommand{\citeppp}[1]{
   \citeauthor{#1} (\citeyear{#1})
}

\newcommand{\R}[1]{}
\newcommand{\lr}[1]{}

\makeatletter
\def \@seccntformat#1{\@ifundefined{#1@cntformat}
{\csname the#1\endcsname\quad}
{\csname #1@cntformat\endcsname}
}
\def\section@cntformat{\thesection.}
\makeatother

\theoremstyle{definition}

\newtheorem{myexmp}{Example}

\newtheorem{lemma}{Lemma}
\newtheorem{thm}{Theorem}

\newcommand{\figref}[1]{\mbox{Figure~\ref{#1}}}
\newcommand{\tabref}[1]{\mbox{Table~\ref{#1}}}
\newcommand{\secref}[1]{\mbox{Section~\ref{#1}}}

\newcommand{\thmref}[1]{\mbox{Theorem~\ref{#1}}}
\newcommand{\lemref}[1]{\mbox{Lemma~\ref{#1}}}

\newcommand{\Tabref}[1]{\mbox{Table~\ref{#1}}}
\newcommand{\Secref}[1]{\mbox{Section~\ref{#1}}}

\begin{document}
\begin{center}
{\Large\bf Orthogonal Gaussian process models}\\
{\Large Matthew Plumlee }\\
\footnotesize {Department of Industrial and Operations Engineering,} \\
{University of Michigan, MI  48109}\\
{(mplumlee@gatech.edu)}\\
\Large  {and V. Roshan Joseph}\\%
\footnotesize {H. Milton Stewart School of Industrial and Systems
Engineering}, \\ {Georgia Institute of Technology, Atlanta, GA 30332}
\end{center}%
\vspace{.25in}
%\linenumbers
\begin{abstract}
Gaussian processes models are widely adopted for nonparameteric/semi-parametric modeling. Identifiability issues occur when the mean model contains polynomials with unknown coefficients. Though resulting prediction is unaffected, this leads to poor estimation of the coefficients in the mean model, and thus the estimated mean model loses interpretability.   This paper introduces a new Gaussian process model whose stochastic part is orthogonal to the mean part to address this issue.  This paper also discusses applications to multi-fidelity simulations using data examples.
\end{abstract}


\newpage
\section{Introduction} \label{sec:intro}
Kriging is a method for estimating or approximating unknown functions from data, with or without noise. For data without noise, it gives an interpolator. A major advantage of kriging over the other interpolators and nonparametric modeling methods is that the prediction intervals can be obtained with almost no additional effort because of its stochastic formulation. Although the method originated  in geostatistics \citepp{matheron1963principles}, kriging has found a prominent place in such diverse fields as uncertainty quantification \citepp{smith2014uncertainty}, spatial statistics \citepp{cressie1993statistics}, computer experiments \citepp{santner2003design}, and machine learning \citepp{rasmussen2006gaussian}, to name a few.

Ordinary kriging, which is arguably the most popular among various kriging methods, can be stated as follows. Consider a deterministic function $f(\cdot)$ that maps the input, $x$ in a bounded subset of $R^d$ labeled $X$, to a scalar valued output, $y(x)$.   An ordinary kriging model considers the family of functions generated by the stochastic process
\begin{equation}
y(x)=\beta+z(x), \label{eq:OK_model}
\end{equation}
where $\beta$ is a mean parameter and $z(x)$ is a zero-mean stochastic process with covariance function $\operatorname{cov} \{z(x),z(x')\} = c(x,x')$. Quite often the stochastic process is assumed to be Gaussian and therefore, the foregoing model can also be called a Gaussian process model. In the Bayesian interpretation, the Gaussian process can be viewed as a prior on the underlying true function $f(\cdot)$ \citepp{currin1991bayesian}.

 This formulation can be broadened by introducing a global trend function $m(x)$ as the mean in the Gaussian process model; this is called universal kriging. The trend function is usually taken as
\[m (x) =  \beta^\mathsf{T} g (x),\]
where $g(x)$ is a vector of  known regression functions and $ \beta = (\beta_1,\ldots,\beta_p)^\mathsf{T}$ is the unknown parameter.

As an example, \cite*{singh2011statistical}  proposed a physically-interpretable regression model for predicting the log-cutting force with respect to depth of cut ($x_1$), cutting speed ($x_2$), laser power ($x_3$), and laser location ($x_4$) in a computer code used for simulating a laser-assisted machining process: $m(x)=\beta_0+\beta_1 \log x_1+\beta_2x_2-\beta_3x_3e^{-\lambda x_4}$. Here, for example, $\exp(-\beta_3)$ could be interpreted as the fraction reduction in the cutting force with a unit increase in the laser power (when $x_4=0$).

The mean function is often left uninterpretable in universal kriging due to an identifiability problem.   For example, consider a model that is equivalent to (\ref{eq:OK_model}),
\begin{equation}
y(x)=\beta+ \bar{z} + z_*(x), \text{ where } \bar{z} = \operatorname{vol} (X)^{-1} \int_X   z(x) dx \text{ and } z_*(x) =  z(x) - \bar{z}. \nonumber
\end{equation}
where $\operatorname{vol} (X)$ is the Lebesgue measure of the set $X$ and thus  $\bar{z}$ represents the  mean of $z(x)$ over the region of interest. Since the integrals are linear operators, we can say that $\bar{z}$ follows a normal distribution and $z_*(\cdot)$ follows a zero-mean Gaussian process. In this formulation it is clear that we have two additive terms in our model that are indistinguishable through the likelihood.

Suppose then we have observed the output at a finite collection of inputs $D = \{x_1,\ldots,x_n\}$, where each $x_i\in X$. Let the observations be $Y := (y_1,\ldots,y_n)^\mathsf{T}$, where $y_i=f(x_i)$. Under the universal kriging model, the best linear unbiased predictor of $y(x)$ is
\[\hat{y}(x) = \hat{\beta}^\mathsf{T} g (x) + c(x, D) C^{-1}\left(Y - G \hat{\beta}\right), \]
where $c(x, D)$ is an $1\times n$ vector with $i$th element $c(x,x_i)$, $C$ is the $n\times n$ covariance matrix with $ij$th element $c(x_i,x_j)$, $G$ is an $n\times p$ model matrix of the regression functions and
\begin{equation}
\hat{\beta} = \left(G^\mathsf{T} C^{-1} G\right)^{-1} G^\mathsf{T} C^{-1} Y. \label{eq:betaUK}
\end{equation}
Here, $\hat{\beta}$ is the generalized least squares estimate of $\beta$ given the assumptions \citepp{kariya2004generalized}.

\begin{figure}
\begin{center}
%\figurebox{}{35pc}{}[examp1.eps]
\includegraphics[width=5in]{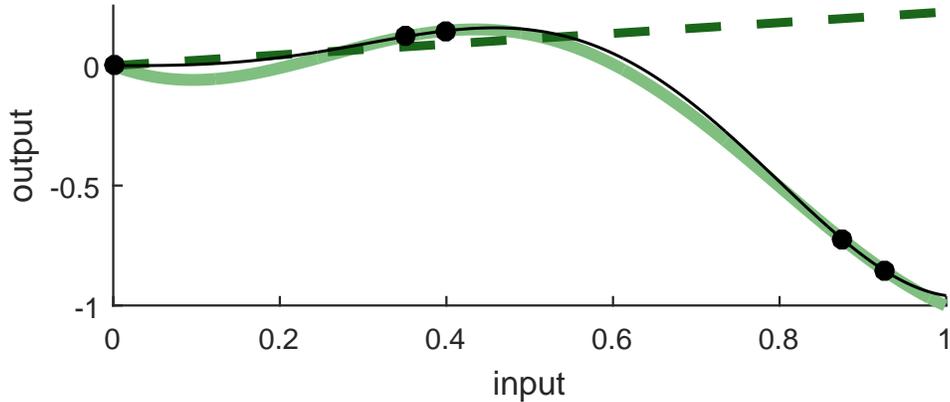}
\caption{ An example of universal kriging discussed in \secref{sec:intro}.  The dots are observations and the thin solid line is the underlying function $x^2 \sin (5 x)$.   The mean is $m(x) = \beta x$.  The thick solid line is the kriging predictor and the thick dashed line is $\hat{m}(x) = \hat{\beta} x$.} \label{fig:examp1}
\end{center}
\end{figure}

As an example, consider the problem of approximating $x^2 \sin(5x)$ in $[0,1]$, as shown in \figref{fig:examp1}, and let $m(x) = \beta x$. The  least squares estimate of $\beta$ is
\[\hat{\beta}^{LS} = \arg \min_\beta \left(Y - G \beta\right)^\mathsf{T} \left(Y - G \beta\right).\]
The value of $\hat{\beta}^{LS}$ in this example is $-0.69$.  The negative sign seems to agree with graph of the function, which on the whole decreases over $[0,1]$.    Now let $c(x,x')=\sigma^2\exp\{- (x-x')^2\}$, where $\sigma^2$ can be any positive constant.
In \figref{fig:examp1}, although the kriging predictor gives a decent approximation on $[0,1]$, the value of $\hat{\beta}$ using (\ref{eq:betaUK}) is $\approx 0.22$.  This appears to imply that as $x$ is increased, the average value of the output increases.  This contradicts the least squares estimate and most statistical intuition. We see that the identifiability problem in the standard universal kriging model can propagate to an unreasonably estimated mean function.

While prediction is still good even with a poor estimation of the mean function, the interpretability of the mean function is lost under the typical, unidentifiable model. This is a disadvantage in  applications such as model calibration \citepp{kennedy2001bayesian} where the mean function contains parameters with physical meaning, and in spatial random effects modeling where the focus is on the estimation of the mean model and not the stochastic term \citepp{reich2006effects}. The identifiability issue has been recognized by many (\citeppp{hodges2010adding}, \citeppp{paciorek2010importance},\citeppp{tuo2015efficient}). We intend to show that the identifiability issue can be overcome by orthogonalizing the Gaussian process term with respect to the mean function.

The idea of making the random field orthogonal to the mean function to avoid identifiability problems was proposed in \cite{reich2006effects} in the context of spatial random effects modeling. \cite{hodges2010adding} used $m(x)=\beta_0+\beta_1x$, for modeling stomach cancer incidence ratio in Solvenia with respect to the socioeconomic scores ($x$). Their approach achieves orthogonality only at the observed locations, which induces two negatives: (i) the stochastic model has a dependency on the observation locations and (ii) outside of the observed locations, such as prediction points, there is no orthogonality.   Extensions of this work in the spatial statistics literature include \cite{hughes2013dimension} and \cite{hanks2015restricted}.  \cite{hanks2015restricted} proposed to make the random field orthogonal to the fixed effects over the entire region $X$, using ``conditioning by Kriging''  \citepp{rue2005gaussian}. Ultimately, they were forced to resort to approximation methods or requiring the orthogonality condition to be met only at the observed locations. We show that by carefully choosing a covariance function we can make the random field orthogonal to the mean function over the entire region $X$. The proposed orthogonal Gaussian process is orthogonal to the mean function.  Our approach is computationally tractable and is amenable to both the frequentist and Bayesian frameworks.

\section{Orthogonal Gaussian process models} \label{sec:OGP}
This section discusses the specification of an orthogonal Gaussian process $z_*(\cdot)$ in general.  The procedure is as follows: given a covariance function $c(\cdot,\cdot)$, replace it with the covariance conditioned on $\int_X g(x) z(x)  d x = 0$, termed $c_*(\cdot,\cdot)$.  A process generated by a zero mean Gaussian process with covariance $c_*(\cdot,\cdot)$ will thus be orthogonal to the mean function.
\subsection{Orthogonalization of a Gaussian Process}
Consider the model
\[y(x) = m(x) + z_*(x),\]
where $z_*(\cdot)$ is a zero mean Gaussian process with covariance function $c_* (\cdot,\cdot)$. $z_*(\cdot)$ is an orthogonal Gaussian process if
 \begin{equation}\label{eq:orthogonality}
 \int_{X} m(\xi) z_*(\xi) \mathrm{d} \xi=0,
 \end{equation}
with probability one. Theorem 1 will show how to construct such an orthogonal Gaussian process when $m(x)=\beta^\mathsf{T} g (x)$. Here we assume $g(x)$ to be a known function. Let $c(\cdot,\cdot)$ be a valid covariance function on $X \times X$ and let
\begin{align}
h(x ) = \int_{X} c(x,\xi ) g(\xi)  \mathrm{d} \xi \text{ and } H = \int_{X} \int_{X} c(\xi',\xi) g(\xi) g(\xi')^\mathsf{T} \mathrm{d} \xi \mathrm{d} \xi'  . \label{eq:integral}
\end{align}
\begin{lemma}\label{lem:PD}
Suppose $X$ is a bounded subset of $R^d$,  $g(\cdot)$ is bounded on $X$, $c(\cdot,\cdot)$  is bounded and continuous on $X \times X$, and $H$ is positive definite, then
$c_*(x,x') =  c(x, x') - h(x)^\mathsf{T} H^{-1} h(x')$
is a semi-positive definite function on $X \times X$.
\end{lemma}
If the covariance function  $c_*(\cdot,\cdot)$ is used to define a Gaussian process $z_*(\cdot)$, then it will meet the orthogonality criteria.
\begin{thm}\label{thm:zero_correlation}
Suppose $X$ is a bounded subset of $R^d$,  $g(\cdot)$ is bounded on $X$, $c(\cdot,\cdot)$  is bounded and continuous on $X \times X$, $H$ is finite and positive definite, and $\beta$ is finite. Then $z_*(\cdot)$ is an orthogonal Gaussian process if
\begin{align}
c_*(x,x') = & c(x, x') - h(x)^\mathsf{T} H^{-1} h(x').
\end{align}
\end{thm}
The proof is given in the Appendix. The conditions of the Theorem 1 are sufficient, not necessary.   They allow for a straightforward proof, are easy to verify, and cover the majority of implementations.

With this stochastic process defined, the best linear unbiased predictor  is
\[\hat{y}(x) = \hat{\beta}^\mathsf{T} g (x) + c_*(x, D)C_*^{-1} \left(Y - G \hat{\beta}\right) \]
where
\begin{equation}
\hat{\beta} = \left(G^\mathsf{T} C_*^{-1} G\right)^{-1} G^\mathsf{T} C_*^{-1} Y, \label{eq:betaOGP}
\end{equation}
and $C_*$ is the same as $C$ with the function $c_*(\cdot,\cdot)$ replacing $c(\cdot,\cdot)$.

\subsection{Some Properties}
The models used in universal kriging and the orthogonal Gaussian process have same distribution, with the exception of functionals that have correlation with $g(\cdot)$.     From the proof of Theorem \ref{thm:zero_correlation}, $z_*(\cdot)$ has the same distribution as $z(\cdot) - h(\cdot)^\mathsf{T} H^{-1} \int_X g(\xi) z(\xi) \mathrm{d} \xi$.  If $f(\cdot)$ is such that \[\int_X \int_X f(\xi) g(\xi') c(\xi,\xi') \mathrm{d} \xi \mathrm{d} \xi' = 0,\]
then $\int_X  f(\xi) z(\xi) \mathrm{d} \xi$ has the same distribution as $\int_X f(\xi) z_*(\xi) \mathrm{d} \xi.$
Thus evaluation functionals that are uncorrelated with $g(\cdot)$ have the same distribution as the original process $z(\cdot)$ under the new model $z_*(\cdot)$.

\begin{figure}
\begin{center}
%\figurebox{0.01pc}{35pc}{}[]
\includegraphics[width=6.5 in]{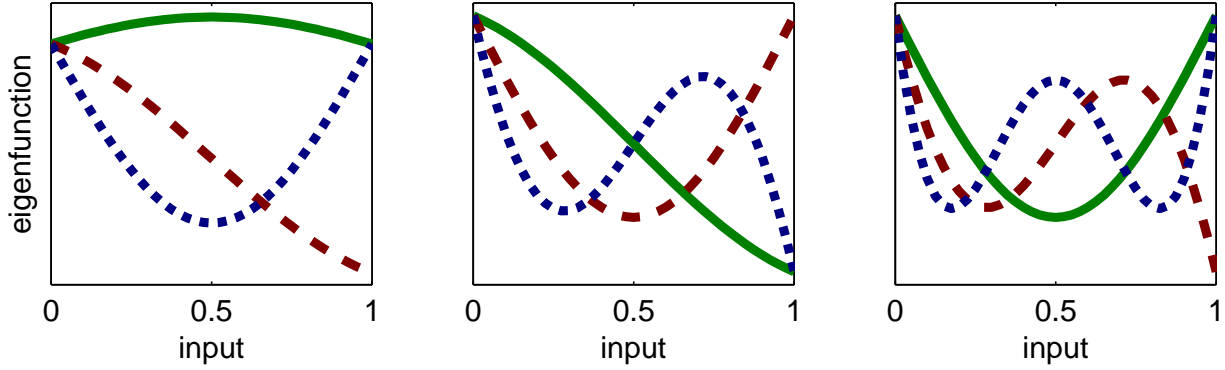}
\caption{Plot of the first three (solid, long dashes, short dashes) eigenfunctions of $z_*(\cdot)$ when $c(x,x') = \exp\left\{-(x-x')^2\right\}$.  The basis functions are none (left),  $g(x) = 1$ (middle), $g(x) = (1,x)^\mathsf{T}$ (right).} \label{fig:cov_plot}
\end{center}
\end{figure}

Consider the eigenfunctions of the orthogonal Gaussian process.  Let $f_{1}(\cdot),\ldots,f_{k}(\cdot),\ldots$ be the orthonormal basis of the corresponding to the Karhunen-Lo\`{e}ve theorem, as in \cite{steinberg2004data}, of $z_*(\cdot)$.  Then
\[z_*(x) =\sum_{k=1}^\infty a_k f_k(x), \]
where $a_k = \int_X f_k(\xi) z_*(\xi) \mathrm{d} \xi$.  These are ordered such that the variance of $a_k$ is decreasing in $k$. The first three eigenfunctions of $z(x)$ with $c(x,x') = \exp\left\{-(x-x')^2\right\}$  are shown in the left panel of \figref{fig:cov_plot}. They look like constant, linear, and quadratic (see also \cite{bursztyn2006comparison}). This could be the reason behind the non-identifiability in the Gaussian process model when the mean function contains lower order polynomials.  Now consider the orthogonal Gaussian process model. The first three eigenfunctions of $z_*(\cdot)$ for the ordinary kriging model are shown in the middle panel of \figref{fig:cov_plot}. They appear linear, quadratic, and cubic.  Thus, the eigenfunction that was close to a constant is now removed from the top three, which suggests less of an identifiability issue if we use a constant in the mean function of the orthogonal Gaussian process model. The right panel of  \figref{fig:cov_plot} shows the first three eigenfunctions of $z_*(\cdot)$ in the universal kriging model with $m(x)=\beta_1+\beta_2x$. As expected, now the constant- and linear-like eigenfunctions are no longer in the top three eigenfunctions and thus do not contribute as much to the distribution of $z_*(\cdot)$.

For ordinary kriging,
\begin{align}
c_*(x,x') = & c(x, x') - \frac{\int_{X} c(x,\xi )  \mathrm{d} \xi \int_{X} c(x',\xi )  \mathrm{d} \xi}{\int_{X} \int_{X} c(\xi,\xi')  \mathrm{d} \xi \mathrm{d} \xi' } . \nonumber
\end{align}
The value of $c_*(x,x)$, the variance of $z_*(x)$, is given by
\begin{align}
c_*(x,x) = & c(x, x) - \text{Constant} \cdot \left\{\int_{X} c(x,\xi )  \mathrm{d} \xi \right\}^2 , \nonumber
\end{align}
which implies different points in the space have differing variances.  Typically points in the middle of $X$, which have significant amounts of integrated covariance, have smaller variance under this model.  To illustrate take $X = [0,1]$ and let $c(x,x') = \exp\{-(x-x')^2/\psi^2\}$, where $\psi$ is a lengthscale parameter, then
\[\int_{\xi \in X} c(x,\xi )  \mathrm{d} \xi \propto   1-\Phi \left(-\sqrt{2} \frac{x}{\psi}\right)-\Phi \left(-\sqrt{2} \frac{1-x}{\psi}\right),\]
where $\Phi$ is the cumulative probability distribution for a Gaussian random variable.  This property carries over to higher dimensional functions.  For example, if the input space is $[0,1]^d$ and the correlation function is the product of one-dimensional Gaussian covariance functions, then the point with the smallest variance of $z_*(\cdot)$ is be the center of the space, $(0.5,0.5,\ldots,0.5)$.

There are implications, using this orthogonal Gaussian process model, in the design of computer experiments. The mean squared prediction error over the space $X$ is
\[\text{MSPE} := \operatorname{vol} (X)^{-1}  \int_{X} \left(\hat{y}(x) - y(x) \right)^2 \mathrm{d} x.\]
The idea is to choose a design $D$  such that $\hat{y}(x)$ that minimizes some aspect of the MSPE, such as its expected value \citepp{sacks1989design}.  Using the orthogonal Gaussian process model, we have that
\[\hat{y} (x)  =g(x)^\mathsf{T} \hat{\beta} +  c_*(x, D)C_*^{-1} \left(Y - G \hat{\beta}\right) = \hat{m}(x) + \hat{z}_* (x) ,\]
where $ \hat{m}(x)$ and  $\hat{z}_* (x) $ are the portions of the predictor that correspond to the mean and random field elements.  Under the conditions of \thmref{thm:zero_correlation},
\begin{align}
\int_{X} \left\{\hat{y}(x) - y(x) \right\}^2 \mathrm{d} x =&  \int_{X} \left\{\hat{z}_*(x) - z_*(x) + \hat{m}(x) - m(x) \right\}^2 \mathrm{d} x  , \nonumber\\
 =& \int_{X} \left\{\hat{m}(x) - m(x) \right\}^2 \mathrm{d} x  + \int_{X} \left\{\hat{z}_*(x) - z_*(x) \right\}^2 \mathrm{d} x  \nonumber\\
  &+2 (\hat{\beta} - \beta)^\mathsf{T} \int_{X} (\hat{z}_*(x) -z_*(x)) g(x) \mathrm{d} x , \nonumber\\
 =&   \int_{X} \left\{\hat{m}(x) - m(x) \right\}^2 \mathrm{d} x  + \int_{X} \left\{\hat{z}_*(x) - z_*(x) \right\}^2 \mathrm{d} x  .\nonumber
\end{align}
Thus, choosing designed experiments under the orthogonal Gaussian process models may be more transparent compared to other universal kriging models.  When using orthogonal Gaussian process models, separate criteria can be used to evaluate the performance of designs for the mean and random field portions of the response.  A thorough investigation of experimental designs under the orthogonal Gaussian process model is outside the scope of this paper.
\section{Fast computation of the covariance function} \label{sec:fast}
After obtaining a predictor, it is evaluated numerous times for optimization and uncertainty quantification. If the integrals in (\ref{eq:integral}) need to be evaluated for each new prediction, the predictor becomes expensive to evaluate and the advantages it possess in terms of estimation and interpretability significantly diminish. It is important to find ways to evaluate the integrals quickly.

For notational simplicity in this section, each vector $x \in X$ has $d$ components labeled $x_1,\ldots, x_d$ and subscripts give dimensional indexing.  Assume the input is located in the Cartesian product of intervals, $X = [a_1,b_1] \times [a_2,b_2]\times\cdots\times[a_d,b_d]$.   Let the function $c(\cdot,\cdot)$ be a separable covariance function, $c(x,x') = \prod_{j=1}^d c_j \left(x_j,x_j' \right),$ where the $c_j$ are positive definite covariance functions on $[a_j,b_j],$ and let $g(x)$ = $g_1 \left(x_1 \right) \odot$ $g_2 \left(x_2\right) \odot$ $\cdots \odot g_d \left(x_d\right),$
where the $g_j(\cdot)$ are mappings from $[a_j,b_j]$ to a $p$-dimensional vector and $\odot$ indicates the Hadamard product.

Here, the high-dimensional integrals can be reduced to single dimensional integrals, i.e.
\begin{align}
 h(x) =& h_1\left(x_1\right)\odot h_2\left(x_2\right) \odot \cdots \odot h_d\left(x_d \right), \quad h_j(x) := \int_{a_j}^{b_j} c_j (x,\xi ) g_j(\xi)  \mathrm{d} \xi, \nonumber\\
H =& H_1 \odot H_2 \odot \cdots \odot H_d, \quad  H_j :=  \int_{a_j}^{b_j}  \int_{a_j}^{b_j} c_j (\xi',\xi) g_j(\xi)  g_j(\xi')^\mathsf{T}  \mathrm{d} \xi \mathrm{d} \xi'. \nonumber
\end{align}

Now consider that each of the basis functions in $g (\cdot)$ are products of one-dimensional linear functions.  Let $i$th element of $g(x)$ be $\prod_{j \in \mathcal{J}_i} x_j,$ where $\mathcal{J}_i$ is subset of $\{1,\ldots,d\}$.  For example, if $\mathcal{J}_i$ is the empty set, then the $i$th element of $g(x)$ is $1$, the constant function and if $\mathcal{J}_i = \{j\}$, then the $i$th element of $g(x)$ is $x_j$.  These basis functions are typically chosen based on effect heredity principle \citepp{wu2009experiments}, under which $\mathcal{J}_i$ is included only if at least one of its subsets defines another basis function (weak heredity) or only if all of its subsets define other basis functions (strong heredity).

Suppose the sets $\mathcal{J}_i$, $i = 1,\ldots,p$, define $g(\cdot)$, and consider the functions
\[\operatorname{M}_j (x) := \int_{a_j}^{b_j} c_j (x,\xi )  \mathrm{d} \xi  , \text{ and } \operatorname{L}_j (x) := \int_{a_j}^{b_j} \xi c_j (x,\xi )  \mathrm{d} \xi ,\]
the mean and $\operatorname{L}$ linear effects, respectively.   The $i$th element of $h(x)$ is $\prod_{j \in \mathcal{J}_i} \operatorname{L}_j (x_j) \prod_{j \notin \mathcal{J}_i} \operatorname{M}_j (x_j).$ If we take
\begin{align}
\operatorname{IM}_j  :=& \int_{a_j}^{b_j} \int_{a_j}^{b_j}   c_j (\xi',\xi )  \mathrm{d} \xi \mathrm{d} \xi'  , \quad \operatorname{IL}_j  := \int_{a_j}^{b_j}  \int_{a_j}^{b_j}  \xi c_j (\xi',\xi )  \mathrm{d} \xi \mathrm{d} \xi' ,  \nonumber\\
\text{ and }& \operatorname{ILL}_j := \int_{a_j}^{b_j}  \int_{a_j}^{b_j} \xi' \xi c_j (\xi',\xi )  \mathrm{d} \xi \mathrm{d} \xi' , \nonumber
\end{align}
 $\operatorname{IM}$ is the integrated mean effect, $\operatorname{IL}$ is the integrated linear effect, and $\operatorname{ILL}$ is the integrated linear-linear effect.   The $ik$th element of the matrix $H$ is then
\[\begin{cases}
\prod_{j \in \mathcal{J}_i \cup \mathcal{J}_k} \operatorname{IL}_j \prod_{j \notin \mathcal{J}_i \cup \mathcal{J}_k } \operatorname{IM}_j,& \text{if } i \neq k \\
\prod_{j \in \mathcal{J}_i} \operatorname{ILL}_j \prod_{j \notin \mathcal{J}_i} \operatorname{IM}_j ,& \text{otherwise}.   \end{cases} \]

Despite the friendly structure, the results still require the evaluation of an integral to find the covariance.  There are a few covariance functions for which these integrals have closed forms.  Thus if
\begin{equation}
c(x,x') = \sigma^2 \exp\left\{-\sum_{j=1}^d \frac{(x_j-x_j')^2}{\psi_j^2} \right\} \label{eq:corr_Gaussian}
\end{equation}
and  $X = [-1,1]^d$, then
\begin{align}
\operatorname{M}_j (x) =&  \frac{\sqrt{\pi}\,\sigma^2  \psi_j}{2}\, \left\{\mathop{\mathrm{erf}}\nolimits\!\left(\frac{x+1}{\psi_j}\right)- \mathop{\mathrm{erf}}\nolimits\!\left(\frac{x-1}{\psi_j}\right)\right\}, \nonumber\\
\operatorname{L}_j (x)  =& \frac{\sigma^2 {\psi_j}^2}{2} \,\left\{ \exp\left({-\frac{(x+1)^2}{{\psi_j}^2}}\right)-\exp\left({-\frac{(x-1)^2}{{\psi_j}^2}}\right)\right\}+ x \operatorname{M}_j (x), \nonumber\\
\operatorname{IM}_j =& 2 \sqrt{\pi}\, \sigma^2 \psi_j\, \mathop{\mathrm{erf}}\nolimits\!\left(\frac{2}{\psi_j}\right)- \sigma^2 {\psi_j}^2\, \left\{1-\exp\left({-\frac{4}{{\psi_j}^2}}\right)\right\}, \quad  \operatorname{IL}_j = 0, \text{ and }  \nonumber\\
\operatorname{ILL}_j =&  \frac{\sigma^2 {\psi_j}^4}{6} \, \left\{ 1-\exp\left({-\frac{4}{{\psi_j}^2}}\right)\right\}   - \frac{\sigma^2 {\psi_j}^2}{3}\, \left\{ 3- \exp\left({-\frac{4}{{\psi_j}^2}}\right)\right\} + \frac{2 \sqrt{\pi}\, \sigma^2 \psi_j}{3}\, \mathop{\mathrm{erf}}\nolimits\!\left(\frac{2}{\psi_j}\right). \nonumber
\end{align}
Here $ \operatorname{IL}_j$ is $0$ because the variables are scaled in $[-1,1]^d$, making $H$ a diagonal matrix, which further simplifies the computation of the covariance function as
\begin{align}
c_*(x,x') = c(x,x') - \sigma^2 \sum_{i=1}^p \prod_{j \in \mathcal{J}_i} \frac{\operatorname{L}_j (x_j) \operatorname{L}_j (x_j')}{ \operatorname{ILL}_j} \prod_{j \notin \mathcal{J}_i} \frac{\operatorname{M}_j (x_j) \operatorname{M}_j (x_j')}{\operatorname{IM}_j } . \label{eq:shortcut}
\end{align}
Similar results can be found for the exponential and the Mat\`{e}rn covariance functions for certain values of the smoothness parameter. The appendix lists two forms of the covariance corresponding to these cases.
\section{Numerical illustrations}
Here, we compare the proposed method, universal kriging,  and another common method using least squares.  In the last method,  we set $\hat{\beta} =\left(G^\mathsf{T} G\right)^{-1} G^\mathsf{T} Y$ and use kriging on the residual, giving
\[\hat{y}(x) = \hat{\beta}^\mathsf{T} g(x) + c(x_0, D)C^{-1} \left(Y - G \hat{\beta}\right).\]
The third method is included because it does not have an obvious identifiability problem and is thus a reasonable comparison.  It produces the same results for the mean function as restricted spatial regression \citepp{hodges2010adding}.
\subsection{One-dimensional input} \label{sec:sub:examp1}
Let $y(x) = \sin (2 x)$ and $X = [0,1]$.    As $x$ is increased, the output increases, thus we try a trend function $m(x) = \beta_1 + \beta_2 x$.  We use three covariance functions to illustrate the ideas (the Gaussian, exponential and the Mat\'{e}rn):
$c(x,x') = \sigma^2 \exp\{-4 (x-x')^2\}, $ $c(x,x') = \sigma^2 \exp\{-2 |x-x'|\}$, and $c(x,x') = \sigma^2 (1+2 |x-x'|) \exp(-2 |x-x'|).$
These are covariances for which we have derived explicit statements for the covariance.  The lengthscale parameter was fixed at a reasonable value for illustrative purposes here. It will be estimated in the next section.
\begin{figure}
\begin{center}
%\figurebox{}{35pc}{}[]
\includegraphics[width=6.5 in]{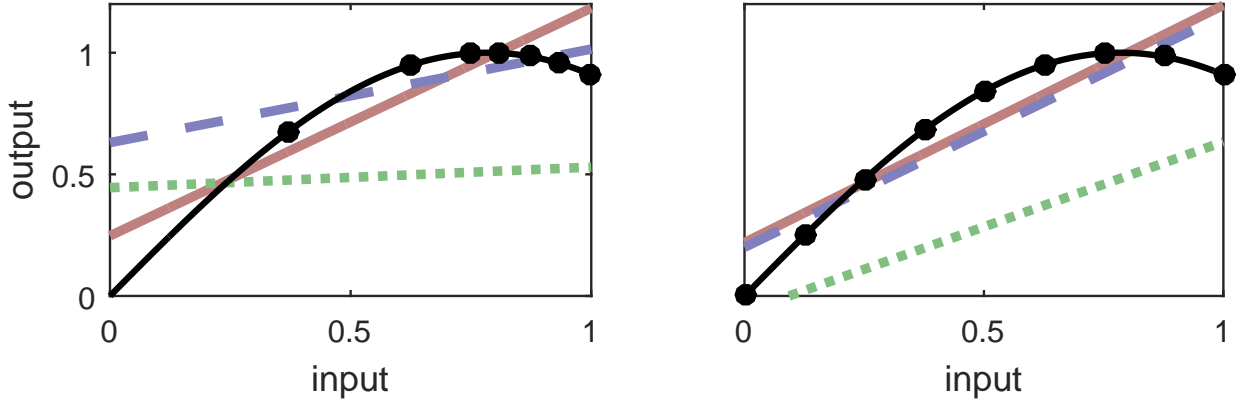}
\caption{ Plot from numerical example in \secref{sec:sub:examp1}. The thick lines are $\hat{m}(x) = \hat{\beta}_1 + \hat{\beta}_2 x$ with orthogonal Gaussian process models (solid line), universal kriging (short dashed line) and least squares (long dashed line) from the numerical example.  The left panel and right panel correspond to different observation schemes.   } \label{fig:examp2}
\end{center}
\end{figure}

We chose two observation schemes: $D = \{0.3725, 0.6225, 0.7475, 0.8100, 0.8725, 0.9350, 0.9975\}$ and $D = \{0, 1/8,\ldots,7/8,1\}$.  \figref{fig:examp2} shows the estimated mean function $\hat{m}(\cdot)$ for all methods under the two observation schemes when the correlation function is Gaussian.    In the first  scheme, the orthogonal Gaussian process method clearly outperforms universal kriging in estimation of the trend; in the second, more friendly observation scheme, universal kriging still does a poor job of estimating the trend due to identifiability issues.

Prediction performance is shown in \tabref{tab:examp1}. The Gaussian has the best predictive power here, but under all choices of covariance functions, the difference between the methods is small.  The major gains of the proposed approach are seen in the estimated parameters, where the mean function parameters are most stably estimated using it when the observation schemes are changed (see the last column of Table 1).
 \begin{table}[t]
\centering
\begin{tabular}{l|rcc|rcc|cc}
    & \multicolumn{3}{c}{Observation Scheme \#1}   & \multicolumn{3}{c}{Observation Scheme\#2}   & &   \\ \hline
    & RMSPE & $\hat{\beta}_1$ & $\hat{\beta}_2$ & RMSPE    & $\hat{\beta}_1$ & $\hat{\beta}_2$ &  $\left|\Delta \hat{\beta}_1\right|$  &   $\left|\Delta \hat{\beta}_2\right|$\\ \hline
LS, Gaussian & $0.40 (10^{-1})$ & $0.63$ & $0.38$ & $5.25 (10^{-5})$ & $0.20$ & $0.95$ & $0.43$ & $0.57$ \\
OGP, Gaussian & $0.20 (10^{-1})$ & $0.25$ & $0.94$ & $5.40 (10^{-5})$ & $0.22$ & $0.98$ & $0.03$ & $0.04$        \\
UK, Gaussian & $0.43 (10^{-1})$ & $0.45$ & $0.08$ & $3.40 (10^{-5})$ & $-0.07$ & $0.70$ & $0.51$ & $0.61$ \\ \hline
OGP, Mat\'{e}rn & $1.30 (10^{-1})$ & $0.43$ & $0.68$ & $222.73 (10^{-5})$ & $0.22$ & $0.97$ & $0.21$ & $0.30$\\
UK,  Mat\'{e}rn  & $1.28 (10^{-1})$ & $0.46$ & $0.29$ & $181.06 (10^{-5})$ & $-0.07$ & $0.82$ & $0.53$ & $0.54$ \\ \hline
OGP, Exponential  & $1.96 (10^{-1})$ & $0.55$ & $0.51$ & $601.44 (10^{-5})$ & $0.22$ & $0.97$ & $0.33$ & $0.46$ \\
UK, Exponential  & $1.91 (10^{-1})$ & $0.58$ & $0.38$ & $519.94 (10^{-5})$ & $0.12$ & $0.92$ & $0.46$ & $0.54$ \\  \hline
\end{tabular}
\caption {Table of the results from the numerical example in \secref{sec:sub:examp1}.  The RMSPE columns are the root mean squared prediction error where the mean squared prediction error is approximated with $400$ equally spaced points.  The columns labeled $\left|\Delta \hat{\beta}_i\right|$ is the absolute value of the difference between $\hat{\beta}_i$s between the two observation schemes. } \label{tab:examp1}
\end{table}
\subsection{Borehole function} \label{sec:sub:examp2}
Consider the borehole function \citepp{worley1987deterministic}, given by
\[y(x) = 2 \pi x_3 (x_4 - x_6) \left\{\log \left(\frac{x_2}{x_1}\right) \left(1+2 \frac{x_3 x_7 }{\log (x_2/x_1) x_1^2 x_8} + \frac{x_3}{x_5} \right)\right\}^{-1} .\]
The space $X$ is the rectangular region $[0.05, 0.15] \times [100, 5000] \times [63070,115600] \times [990,1110]  \times [63.1, 116] \times [700,820] \times [1120,1680] \times [9855,12045]$.  We scaled it $[-1,1]^8$ to use  (\ref{eq:shortcut}).

We took the mean function $m(x) = \beta_1 + \beta_2 x_1 +\ldots \beta_9 x_8$.
We generated a random Latin hypercube design $50$ times for sample sizes of $20$, $40$, $80$ and $160$.  Because of the function's smoothness, we adopted the Gaussian covariance function  (\ref{eq:corr_Gaussian}). We did not a priori chose $\psi = (\psi_1,\ldots,\psi_8)$; we selected it as the maximum likelihood estimate,
\[\operatorname{argmin}_\psi \log   \left[\frac{1}{n}\left\{ Y- G \hat{\beta}(\psi)\right\}^\mathsf{T} C(\psi)^{-1} \left\{ Y- G \hat{\beta}(\psi)\right\}\right]+\frac{1}{n} \log\left\{ \det C(\psi)\right\} ,\]
with $C_*$ replacing $C$ for the orthogonal Gaussian process method and the least squares estimate instead of $\hat{\beta}(\psi)$ for the final method. The above optimization problem was restricted to $[0.1,5]^8$.

The results of the simulations are summarized in the supplementary material. The mean square prediction error is essentially the same under each approach. Figure 4 shows boxplots across 50 replicates for the first two $\hat{\beta}_i$'s.  Using universal kriging, $\hat{\beta}_2$ has only a small reduction in standard deviation as the sample size is increased to $160$, indicating a lack of convergence.    When $n=160$, the standard deviation of the estimate of $\hat{\beta}_2$ using orthogonal Gaussian processes is ten times smaller than the estimate using universal kriging.   The results using universal kriging are significantly worse for $\hat{\beta}_1$, which appears to be centered at different values depending on the sample size.  While the least squares estimates appear to be converging to reasonable values,  convergence is slow relative to the orthogonal Gaussian process model.  From a modeling standpoint, this may be because the least squares estimates do not incorporate the smoothness of the response \citepp{tuo2015efficient}.
\begin{figure}
\begin{center}
%\figurebox{}{35pc}{}[]
\includegraphics[width=5in]{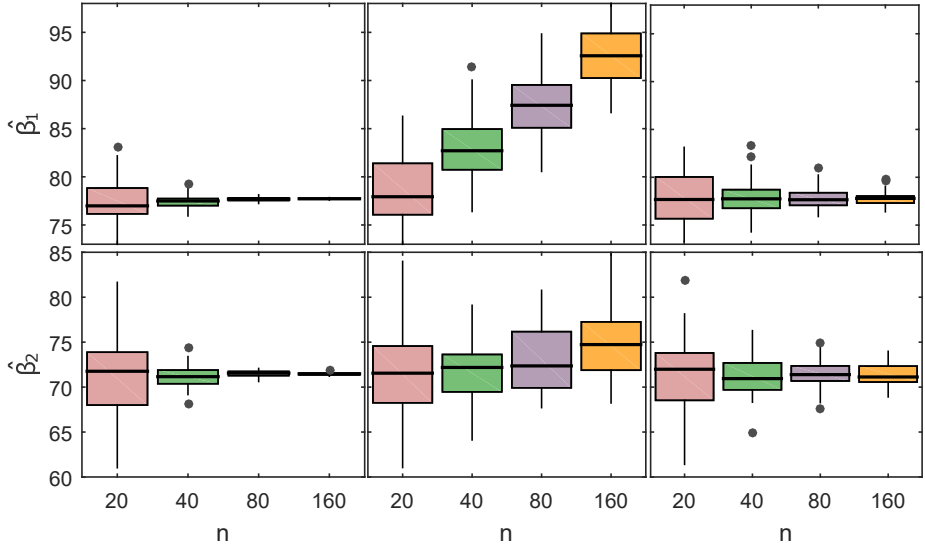}
\caption{ Boxplots from the numerical example in \secref{sec:sub:examp2} of $\hat{\beta}_1$ and $\hat{\beta}_2$ from the 50 replications using the orthogonal Gaussian process (left), universal kriging (middle) and least squares (right).} \label{fig:examp4}
\end{center}
\end{figure}

\begin{figure}
\begin{center}
%\figurebox{}{35pc}{}[]
\includegraphics[width=6.5 in]{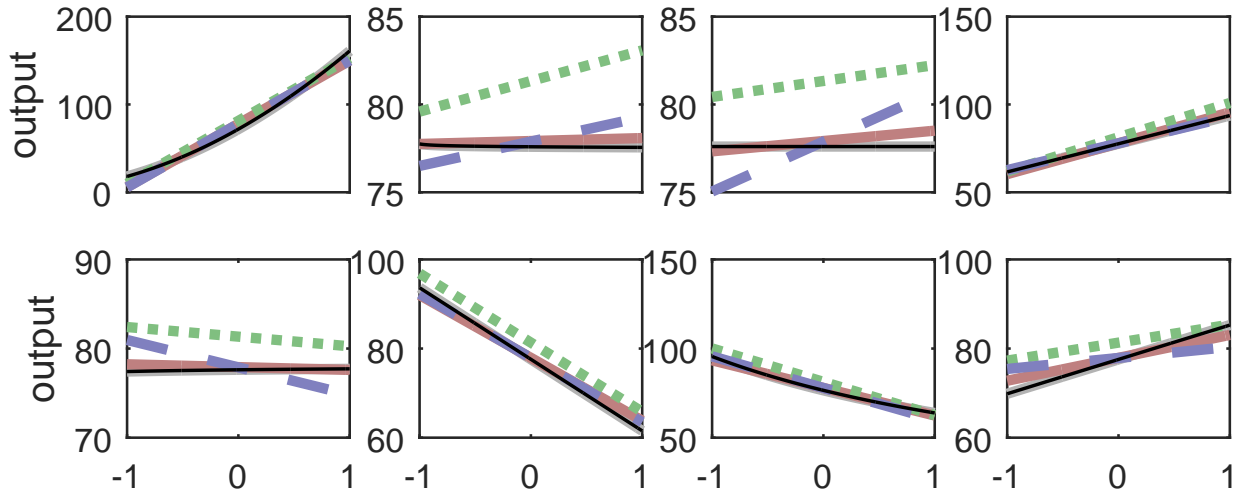}
\caption{ A plot from the numerical example in \secref{sec:sub:examp2} when $n=40$.  Each plot is of functions with  all but the $i$th variable integrated out, where on the top row $i=1,2,3,4$ and on the bottom row $i=5,6,7,8$.  The true function is in the thin dark line.  The thick lines are  $\hat{m}(x)$ using orthogonal Gaussian process models (solid line), universal kriging (short dashed line) and least squares (long dashed line).  All inputs are scaled to the domain $[-1,1]^8$ for comparison.   } \label{fig:examp3}
\end{center}
\end{figure}

\figref{fig:examp3} contains an illustration when $n=40$.  To show the high-dimensional functions, we plot the functions while integrating all but the $i$th variable, $i = 1,\ldots, 8$.   Here, a user using $\hat{m}(x)$ to understand the borehole function may reach incorrect conclusions by choosing universal kriging or least squares over the orthogonal Gaussian process method.  This can have important implications in identifying the causes of discrepancy in model calibration problems \citepp{joseph2015engineering}.

\section{Application in multi-fidelity simulations}  \label{sec:approx}
\R{AE11}
A frequently encountered problem in computer experiment analysis is the fusion of multi-fidelity simulations  (\citeppp{kennedy2000predicting},\citeppp{tuo2014surrogate}). For example, consider simulations with only two levels of accuracy (\citeppp{qian2006building},\citeppp{qian2008bayesian}). The low-accuracy simulations are cheaper, whereas the high-accuracy simulations are expensive and the aim is to predict the output using fewer expensive simulations. Following \cite{kennedy2000predicting},  let $g(x)$ be $(1,y_0(x))^\mathsf{T}$, where $y_0(x)$ is the low-accuracy response and $y(x)$ is the high-accuracy response.   The model is thus $y(x)=\beta_1 + \beta_2 y_0(x) +z(x), $ where $z(x)$ is a Gaussian process. For simplicity, we assume $y_0(x)$ to be available in analytical form.  Since $\beta_1$ and $\beta_2$ have particular meaning in this context, it becomes imperative that they be estimated well, and the universal kriging approach (which is implied by \cite{kennedy2000predicting}) can give misleading results. \R{AE15}

To illustrate these ideas, consider two examples presented in \cite{ba2013integrating} and \cite{qian2006building}.  In the first, two responses exist and are analyzed separately.   Since the low-accuracy response used by \cite{qian2006building} was unavailable to us,  the linear low-accuracy response $y_0(x) = -7.97 + (2920, -0.257, 0.0119, 0.266)  x$ was used instead, with $x$ a vector of the design variables as ordered in Table 2 of \cite{qian2006building}.  The $R^2$ between their data and this linear model was over $95 \%$.       The covariance was a Mat\`{e}rn covariance covariance function with the lengthscale parameters equal to the twice the ranges of each input.  For the computation of $h(\cdot)$ and $H$, numerical quadrature was used to solve to sufficient accuracy.   The integration results from \Secref{sec:fast} can be used to simplify the expression for  $h(\cdot)$ and $H$ in the second example because the model is linear. There is no ability to compute RMSPE because the high-accuracy response can only be evaluated a finite number of times due to time and resource constraints, as detailed in the respective papers.

\Tabref{tab:examp3} shows the different conclusions one arrives at when using the estimation methods. In the \cite{ba2013integrating} example, universal kriging produced significantly different parameters from the other methods.  The universal kriging estimates of the location parameters are very large in both examples.  For the second response, the universal kriging estimate of $\hat{\beta}_2$ is negative, indicating that the low-accuracy model does not correspond to the high-accuracy response.  If universal kriging's estimated parameters were directly used to draw conclusions, a user might wish go back and adjust the low-accuracy responses.  The orthogonal approach tells a different story, with  $\hat{\beta}_1$ close to $0$ and $\hat{\beta}_2$ close to $1$ for both responses.  Thus the parameters from the orthogonal approach indicate that the low-accuracy responses are good representations of the high-accuracy responses.    For the \cite{qian2006building} example, all approaches similarly estimate the scale parameter, but the sign of the estimated location parameter is flipped with the orthogonal Gaussian process method.   Moreover, both the orthogonal approach and the universal kriging approach place the scale parameter very close to $1$ compared to the least squares approach.  Thus in both datasets, information gleaned from the  estimated parameters varies based the chosen estimation method.   Based on the simulations in Sections \ref{sec:sub:examp1} and \ref{sec:sub:examp2}, it would appear that the orthogonal Gaussian process estimates have some advantages over the other two approaches.
 \begin{table}[t]%
\centering
\begin{tabular}{l|cc|cc|cc}
   & \multicolumn{4}{c}{\cite{ba2013integrating}}   & \multicolumn{2}{c}{\cite{qian2006building}}        \\ \hline
    &\multicolumn{2}{c}{response $\#$ 1} & \multicolumn{2}{c}{response $\#$ 2}   & \\ \hline
    &   $\hat{\beta}_1$ & $\hat{\beta}_2$ & $\hat{\beta}_1$ & $\hat{\beta}_2$ &   $\hat{\beta}_1$ & $\hat{\beta}_2$ \\ \hline
    LS  & 1.43 & 1.06 & 0.33 & 0.99 & -0.35 & 1.03 \\
OGP  & 0.43 & 1.08 & 1.43 & 0.99 & 0.16 & 1.00 \\
UK  & -34.12 & 1.29 & 88.03 & -1.67 & -0.66 & 1.00
\end{tabular}
\caption {Table of the results from the examples in \secref{sec:approx}.  } \label{tab:examp3}
\end{table}
\R{AE12}

\section{Conclusions and discussion}
The accurate estimation of the mean function is important in many applications involving computer model calibration, global sensitivity analysis, and spatial random effects modeling. The identifiability problem can be avoided by making the mean function orthogonal to the random field, and in this article, we have proposed a Gaussian process model that has this property. This is achieved by modifying a given covariance function to incorporate the orthogonality condition. While these covariances are defined by integrals, some versions of the covariance functions can be quickly evaluated.  This paper has focused on the modeling aspect of the problem and numerical examples show quite promising results for estimation.

\section{Supplementary Materials}
The online Supplementary Materials contains additional information for section 4.2 in the paper.

\section*{Acknowledgements}
This research is supported by a U.S. National Science Foundation grant CMMI-1266025 and a U.S. Department of Energy Advanced Scientific Computing Research Award ERKJ259.
\appendix
\section*{Appendix: Proof of \lemref{lem:PD}}
Since $c(\cdot,\cdot)$ is continuous, we can choose a measurable version of the process $z(\cdot)$  such that
\[E \left\{\int_{\xi \in X} z(\xi)^2 \mathrm{d} \xi \right\} = \int_{\xi \in X} E(z(\xi) ^2) \mathrm{d} \xi  = \int_{\xi \in X} r(\xi,\xi)\mathrm{d} \xi < \infty\]
\citepp{marcus1972sample}. Then
\[E\left\{\int_{\xi \in X} |z(\xi)  g_i(\xi)| \mathrm{d} \xi\right\}  < \infty, E\left\{\int_{\xi \in X}\int_{\xi' \in X} |z(\xi) z(\xi')  g_i (\xi)  g_i (\xi') |\mathrm{d} \xi'  \mathrm{d} \xi \right\} < \infty,  \]
by the assumption of $g(\cdot)$ being bounded.    Thus we can appropriately switch integral and expectation signs and note that the covariance matrix of $\left\{z(x_1),\ldots,z(x_n),\int_{\xi \in X} z(\xi) g (\xi) \mathrm{d} \xi\right\}^\mathsf{T} $ is
\[\left( \begin{array}{ccc}
C & W  \\
W^\mathsf{T} & H \end{array} \right) = \left(\begin{array}{ccc}
I & W H^{-1}  \\
0 & I \end{array}  \right) \left(\begin{array}{ccc}
C - W H^{-1} W^\mathsf{T}  &0  \\
0 & H \end{array}  \right) \left(\begin{array}{ccc}
I & W H^{-1}  \\
0 & I \end{array}  \right)^\mathsf{T}, \]
where $W$ is the $n \times p$ matrix $\left\{h(x_1),\ldots,h(x_n)\right\}.$   Since $c(\cdot,\cdot)$ is positive definite, this matrix is positive semidefinite and $C - W H^{-1} W^\mathsf{T} $ is positive semidefinite.  As $C_* = C - W H^{-1} W^\mathsf{T}$, $C_*$ is positive semidefinite for all $D$.

\section*{Appendix: Proof of \thmref{thm:zero_correlation}}
We have that $m(x) = \beta^\mathsf{T} g (x)$ and then
\[\int_{\xi \in X} m(\xi) z_*(\xi) \mathrm{d} \xi = \int_{\xi \in X} \left\{\beta^\mathsf{T} g (\xi)\right\} z_*(\xi) \mathrm{d} \xi = \beta^\mathsf{T} \left\{\int_{\xi \in X} z_*(\xi) g (\xi) \mathrm{d} \xi\right\}  .\]
Consider the $i$th element of $\int_{\xi \in X} z_*(\xi) g (\xi) \mathrm{d} \xi$, $\int_{\xi \in X} z_*(\xi) g_i (\xi) \mathrm{d} \xi $, where $g_i(x)$ is the $i$th value of $g(x)$.   We show this has mean $0$ and variance $0$, thus is $0$ with probability one.

Since $c_*(\cdot,\cdot)$ is  continuous, we can choose a measurable version of the process $z_*(\cdot)$  and then
\[E \left\{\int_{\xi \in X} z(\xi)^2 \mathrm{d} \xi \right\}  = \int_{\xi \in X} r(\xi,\xi)\mathrm{d} \xi - \int_{\xi \in X} h(\xi)^\mathsf{T} H^{-1} h(\xi) \mathrm{d} \xi  \leq \int_{\xi \in X} r(\xi,\xi)\mathrm{d} \xi < \infty.\]
As in \lemref{lem:PD}, we can switch the expectation and  integral to get
\[E\left\{\int_{\xi \in X} z_*(\xi)  g_i(\xi) \mathrm{d} \xi\right\} =\int_{\xi \in X}  E\left\{z_*(\xi)\right\}  g_i(\xi)  \mathrm{d} \xi = 0,\]
\[
\operatorname{var} \left\{\int_{\xi \in X} z_*(\xi)  g_i(\xi) \mathrm{d} \xi\right\} = \int_{q' \in X}  \int_{q \in X} c_*(q,q')  g_i(q) g_i(q') \mathrm{d} q   \mathrm{d} q' .\]
Expanding the variance term yields
\begin{align}
\int_{q' \in X}  \int_{q \in X} c_*(q,q')   &g_i(q) g_i(q') \mathrm{d} q   \mathrm{d} q' =  \int_{q' \in X}  \int_{q \in X}c(q, q') g_i(q) g_i(q')\mathrm{d} q   \mathrm{d} q' \nonumber\\
- &\left(\int_{q \in X} \int_{\xi \in X} c(q,\xi ) g_i(q) g(\xi)^\mathsf{T}  \mathrm{d} \xi   \mathrm{d} q\right) \left(\int_{\xi \in X}\int_{\xi' \in X} c(\xi,\xi') g(\xi) g(\xi')^\mathsf{T} \mathrm{d} \xi'  \mathrm{d} \xi \right)^{-1} \nonumber\\
&  \cdot \left( \int_{q' \in X} \int_{\xi \in X} c(\xi,q') g_i(q') g(\xi) \mathrm{d} \xi \mathrm{d} q'  \right)  \nonumber\\
=& \int_{q \in X} \int_{q' \in X} c(q, q') g_i(q) g_i(q') \mathrm{d} q   \mathrm{d} q'   - \int_{q' \in X} \int_{\xi \in X} c(\xi,q' )  g_i(\xi) g_i(q') \mathrm{d} \xi \mathrm{d} q' = 0 . \nonumber
\end{align}
\section*{Appendix: Other versions of the orthogonal covariance}
This section details other explicit forms of the covariance that mirror the statement of
(\ref{eq:shortcut}).  Let each vector $x \in X$ have $d$ components labeled $x_1,\ldots, x_d$.   We use subscripts for all dimensional indexing in this section.  Assume the input is located in the Cartesian product of standard intervals, $X = [-1,1]^d$.    Let the $i$th element of $g(x)$ be $\prod_{j \in \mathcal{J}_i} x_j,$ where $\mathcal{J}_i$ is subset of $\{1,\ldots,d\}$. Consider the exponential covariance \[c(x,x') = \sigma^2 \prod_{j=1}^d \exp(-|x_j-x_j'|/\psi_j).\]
 When it is used, the values of the function and constants in (\ref{eq:shortcut}) are
\begin{align}
\operatorname{M}_j (x)   =   &  - \psi_j  \left(\exp \left(\frac{x - 1}{\psi_j}\right)  + \exp \left(-\frac{x + 1}{\psi_j}\right)  - 2\right) ,\nonumber \\
 \operatorname{L}_j (x)   =   &\psi_j  \left(\psi_j + x\right)  - \psi_j  \left(\psi_j - x\right)  - \psi_j  \exp \left(\frac{x - 1}{\psi_j}\right)   \left(\psi_j + 1\right)  \nonumber\\
 & + \psi_j  \exp \left(-\frac{x + 1}{\psi_j}\right)   \left(\psi_j + 1\right), \nonumber \\
 \operatorname{IM}_j(x)   =   &4  \psi_j   + 2  {\psi_j}^2  \left(\exp \left(-\frac{2}{\psi_j}\right)  - 1\right), \nonumber \\
 \operatorname{ILL}_j (x)   =   &\frac{4  \psi_j}{3}  + 2  \psi_j  \left(\psi_j  \left(\psi_j - 1\right) - \psi_j  \exp \left(-\frac{2}{\psi_j}\right)   \left(\psi_j + 1\right)\right)  \nonumber\\
 & + 2  {\psi_j}^2  \left(\psi_j  \left(\psi_j - 1\right) - \psi_j  \exp \left(-\frac{2}{\psi_j}\right)   \left(\psi_j + 1\right)\right).\nonumber
\end{align}

The Mat\`{e}rn covariance, when the smoothness parameter is given by $3/2$,  is
\[c(x,x') = \sigma^2 \prod_{j=1}^d (1+|x_j-x_j'|/\psi_j) \exp(-|x_j-x_j'|/\psi_j).\]
When it is used, the values of the function and constants in (\ref{eq:shortcut}) are
\begin{align}
\operatorname{M}_j (x)   =   &2  \psi_j  - \psi_j  \left(\exp \left(\frac{x - 1}{\psi_j}\right)  + \exp \left(-\frac{x + 1}{\psi_j}\right)  - 2\right)  - \exp \left(-\frac{x + 1}{\psi_j}\right)   \left(\psi_j + x + 1\right)  \nonumber\\
 & - \exp \left(\frac{x - 1}{\psi_j}\right)   \left(\psi_j - x + 1\right), \nonumber \\
 \operatorname{L}_j (x)   =   &\psi_j  \left(\psi_j + x\right)  + 2  \psi_j  x  - \exp \left(\frac{x - 1}{\psi_j}\right)   \left(2  \psi_j - x - \psi_j  x + 2  {\psi_j}^2 + 1\right)  \nonumber\\
 &- \psi_j  \left(\psi_j - x\right)  + \exp \left(-\frac{x + 1}{\psi_j}\right)   \left(2  \psi_j + x + \psi_j  x + 2  {\psi_j}^2 + 1\right)  \nonumber\\
 &- \psi_j  \exp \left(\frac{x - 1}{\psi_j}\right)   \left(\psi_j + 1\right)   + \psi_j  \exp \left(-\frac{x + 1}{\psi_j}\right)   \left(\psi_j + 1\right), \nonumber \\
 \operatorname{IM}_j(x)   =   & 2  \psi_j  \left(2  \exp \left(-\frac{2}{\psi_j}\right)  - 3  \psi_j + 3  \psi_j  \exp \left(-\frac{2}{\psi_j}\right)  + 4\right), \nonumber \\
 \operatorname{ILL}_j (x)   =   &\frac{8  \psi_j}{3}  - 4  \psi_j  \exp \left(-\frac{2}{\psi_j}\right)   - 14  {\psi_j}^2  \exp \left(-\frac{2}{\psi_j}\right)   - 20  {\psi_j}^3  \exp \left(-\frac{2}{\psi_j}\right)   \nonumber\\
 &- 10  {\psi_j}^4  \exp \left(-\frac{2}{\psi_j}\right)   - 6  {\psi_j}^2   + 10  {\psi_j}^4. \nonumber
    \end{align}

\bibliographystyle{rss}
\bibliography{OGPM_ref}

\begin{thebibliography}{28}
\expandafter\ifx\csname natexlab\endcsname\relax\def\natexlab#1{#1}\fi
\expandafter\ifx\csname url\endcsname\relax
  \def\url#1{\texttt{#1}}\fi
\expandafter\ifx\csname urlprefix\endcsname\relax\def\urlprefix{URL: }\fi

\bibitem[{Ba et~al.(2013)Ba, Jain, Joseph and Singh}]{ba2013integrating}
Ba, S., Jain, N., Joseph, V.~R. and Singh, R. (2013) Integrating analytical
  models with finite-element models: an application in micromachining.
\newblock \textit{Journal of Quality Technology}, \textbf{45}, 200--213.

\bibitem[{Bursztyn and Steinberg(2006)}]{bursztyn2006comparison}
Bursztyn, D. and Steinberg, D.~M. (2006) Comparison of designs for computer
  experiments.
\newblock \textit{Journal of Statistical Planning and Inference}, \textbf{136},
  1103--1119.

\bibitem[{Cressie and Cassie(1993)}]{cressie1993statistics}
Cressie, N.~A. and Cassie, N.~A. (1993) \textit{Statistics for {S}patial
  {D}ata}.
\newblock Wiley New York.

\bibitem[{Currin et~al.(1991)Currin, Mitchell, Morris and
  Ylvisaker}]{currin1991bayesian}
Currin, C., Mitchell, T., Morris, M. and Ylvisaker, D. (1991) Bayesian
  prediction of deterministic functions, with applications to the design and
  analysis of computer experiments.
\newblock \textit{Journal of the American Statistical Association},
  \textbf{86}, 953--963.

\bibitem[{Hanks et~al.(2015)Hanks, Schliep, Hooten and
  Hoeting}]{hanks2015restricted}
Hanks, E.~M., Schliep, E.~M., Hooten, M.~B. and Hoeting, J.~A. (2015)
  Restricted spatial regression in practice: geostatistical models,
  confounding, and robustness under model misspecification.
\newblock \textit{Environmetrics}, \textbf{26}, 243--254.

\bibitem[{Hodges and Reich(2010)}]{hodges2010adding}
Hodges, J.~S. and Reich, B.~J. (2010) Adding spatially-correlated errors can
  mess up the fixed effect you love.
\newblock \textit{The American Statistician}, \textbf{64}, 325--334.

\bibitem[{Hughes and Haran(2013)}]{hughes2013dimension}
Hughes, J. and Haran, M. (2013) Dimension reduction and alleviation of
  confounding for spatial generalized linear mixed models.
\newblock \textit{Journal of the Royal Statistical Society: B}, \textbf{75},
  139--159.

\bibitem[{Joseph and Yan(2015)}]{joseph2015engineering}
Joseph, V.~R. and Yan, H. (2015) Engineering-driven statistical adjustment and
  calibration.
\newblock \textit{Technometrics}, \textbf{57}, 257--267.

\bibitem[{Kariya and Kurata(2004)}]{kariya2004generalized}
Kariya, T. and Kurata, H. (2004) \textit{Generalized {L}east {S}quares}.
\newblock John Wiley \& Sons.

\bibitem[{Kennedy and O'Hagan(2000)}]{kennedy2000predicting}
Kennedy, M.~C. and O'Hagan, A. (2000) Predicting the output from a complex
  computer code when fast approximations are available.
\newblock \textit{Biometrika}, \textbf{87}, 1--13.

\bibitem[{Kennedy and O'Hagan(2001)}]{kennedy2001bayesian}
--- (2001) Bayesian calibration of computer models.
\newblock \textit{Journal of the Royal Statistical Society: B}, \textbf{63},
  425--464.

\bibitem[{Marcus and Shepp(1972)}]{marcus1972sample}
Marcus, M.~B. and Shepp, L.~A. (1972) Sample behavior of gaussian processes.
\newblock In \textit{Proceedings of the Sixth Berkeley Symposium on
  Mathematical Statistics and Probability} (eds. L.~M. {Le Cam}, J.~Neyman and
  E.~L. Scott), vol.~2, 423--441. Berkeley, Calif.: University of California
  Press.

\bibitem[{Matheron(1963)}]{matheron1963principles}
Matheron, G. (1963) Principles of geostatistics.
\newblock \textit{Economic Geology}, \textbf{58}, 1246--1266.

\bibitem[{Paciorek(2010)}]{paciorek2010importance}
Paciorek, C.~J. (2010) The importance of scale for spatial-confounding bias and
  precision of spatial regression estimators.
\newblock \textit{Statistical {S}cience}, \textbf{25}, 107--125.

\bibitem[{Qian and Wu(2008)}]{qian2008bayesian}
Qian, P. Z.~G. and Wu, C. F.~J. (2008) Bayesian hierarchical modeling for
  integrating low-accuracy and high-accuracy experiments.
\newblock \textit{Technometrics}, \textbf{50}, 192--204.

\bibitem[{Qian et~al.(2006)Qian, Seepersad, Joseph, Allen and
  Wu}]{qian2006building}
Qian, Z., Seepersad, C.~C., Joseph, V.~R., Allen, J.~K. and Wu, C. F.~J. (2006)
  Building surrogate models based on detailed and approximate simulations.
\newblock \textit{Journal of Mechanical Design}, \textbf{128}, 668--677.

\bibitem[{Rasmussen and Williams(2006)}]{rasmussen2006gaussian}
Rasmussen, C.~E. and Williams, C. (2006) \textit{Gaussian Processes for Machine
  Learning}.
\newblock MIT Press.

\bibitem[{Reich et~al.(2006)Reich, Hodges and Zadnik}]{reich2006effects}
Reich, B.~J., Hodges, J.~S. and Zadnik, V. (2006) Effects of residual smoothing
  on the posterior of the fixed effects in disease-mapping models.
\newblock \textit{Biometrics}, \textbf{62}, 1197--1206.

\bibitem[{Rue and Held(2005)}]{rue2005gaussian}
Rue, H. and Held, L. (2005) \textit{Gaussian Markov Random Fields: Theory and
  {A}pplications}.
\newblock CRC Press.

\bibitem[{Sacks et~al.(1989)Sacks, Welch, Mitchell and Wynn}]{sacks1989design}
Sacks, J., Welch, W.~J., Mitchell, T.~J. and Wynn, H.~P. (1989) Design and
  analysis of computer experiments.
\newblock \textit{Statistical Science}, \textbf{4}, 409--423.

\bibitem[{Santner et~al.(2003)Santner, Williams and Notz}]{santner2003design}
Santner, T.~J., Williams, B.~J. and Notz, W. (2003) \textit{The Design and
  Analysis of Computer Experiments}.
\newblock Springer.

\bibitem[{Singh et~al.(2011)Singh, Joseph and Melkote}]{singh2011statistical}
Singh, R.~K., Joseph, V.~R. and Melkote, S.~N. (2011) A statistical approach to
  the optimization of a laser-assisted micromachining process.
\newblock \textit{The International Journal of Advanced Manufacturing
  Technology}, \textbf{53}, 221--230.

\bibitem[{Smith(2014)}]{smith2014uncertainty}
Smith, R.~C. (2014) \textit{Uncertainty Quantification: Theory, Implementation,
  and Applications}.
\newblock SIAM.

\bibitem[{Steinberg and Bursztyn(2004)}]{steinberg2004data}
Steinberg, D.~M. and Bursztyn, D. (2004) Data analytic tools for understanding
  random field regression models.
\newblock \textit{Technometrics}, \textbf{46}, 411--420.

\bibitem[{Tuo et~al.(2014)Tuo, Wu and Yu}]{tuo2014surrogate}
Tuo, R., Wu, C. F.~J. and Yu, D. (2014) Surrogate modeling of computer
  experiments with different mesh densities.
\newblock \textit{Technometrics}, \textbf{56}, 372--380.

\bibitem[{Tuo et~al.(2015)Tuo, Wu et~al.}]{tuo2015efficient}
Tuo, R., Wu, C.~J. et~al. (2015) Efficient calibration for imperfect computer
  models.
\newblock \textit{The Annals of Statistics}, \textbf{43}, 2331--2352.

\bibitem[{Worley(1987)}]{worley1987deterministic}
Worley, B.~A. (1987) Deterministic uncertainty analysis.
\newblock \textit{Tech. Rep. ORN-0628}, Oak Ridge National Lab., TN (USA).

\bibitem[{Wu and Hamada(2009)}]{wu2009experiments}
Wu, C. F.~J. and Hamada, M.~S. (2009) \textit{Experiments: Planning, Analysis,
  and Optimization}.
\newblock John Wiley \& Sons, 2 edn.

\end{thebibliography}

\end{document}